\documentclass[11pt]{article}

\usepackage[utf8]{inputenc}
\usepackage[T1]{fontenc}

\usepackage{lmodern}
\usepackage{amsmath,amssymb,amsfonts}
\usepackage{graphicx}
\usepackage{booktabs}
\usepackage{multirow}
\usepackage{caption}
\usepackage{subcaption}
\usepackage{xcolor}
\usepackage{float}
\usepackage{url}
\usepackage{hyperref}
\usepackage{ragged2e}
\usepackage{authblk}

\usepackage[a4paper,margin=1in]{geometry}

\hypersetup{
    colorlinks=true,
    linkcolor=blue,
    citecolor=blue,
    urlcolor=blue
}

\title{
A Scalable Digital Twin Framework for Energy Optimization in Data Centers
}

\author{
Raphael Hendrigo de Souza Gonçalves$^{1}$\\
\small Federal University of São João del-Rei (UFSJ)\\
\small \texttt{raphael.goncalves@tcmsp.tc.br}\\
\small ORCID: 0009-0009-3635-6395
\and
Wendel Marcos dos Santos$^{2}$\\
\small Federal Institute of Education, Science and Technology of São Paulo (IFSP)\\
\small \texttt{wendel.santos@ifsp.edu.br}\\
\small ORCID: 0000-0002-1336-3623
}

\date{}

\usepackage{abstract}

\begin{document}
\maketitle

% ------------------------------------------------
% ABSTRACT
% ------------------------------------------------

\begin{abstract}
\noindent
This study proposes a scalable Digital Twin framework for energy optimization in data centers. The framework integrates IoT-based data acquisition, cloud computing, and machine learning techniques to enable real-time monitoring, forecasting, and intelligent energy management. A controlled small-scale data center environment was developed to monitor variables such as power consumption, temperature, and computational workload. Long Short-Term Memory (LSTM) models were employed to predict energy demand and support operational decision-making. Experimental results demonstrated improvements in energy efficiency, including reductions in power consumption and enhancements in Power Usage Effectiveness (PUE). Despite being evaluated in a constrained environment, the proposed framework demonstrates strong potential as a scalable and cost-effective solution for sustainable data center management.

\end{abstract}

%\vspace{0.5em}


% ------------------------------------------------
% SECTIONS
% ------------------------------------------------

\section{Introduction}

The growing demand for cloud computing, big data, artificial intelligence, Internet of Things (IoT), and the continuous availability of critical applications has positioned data centers as central elements in modern technological infrastructure \cite{alam2023, barroso2009}. According to \cite{koomey2011} and recent projections from the International Energy Agency, energy consumption associated with these infrastructures has increased significantly over the past decades, driven by the expansion of digital services such as streaming platforms, social networks, e-commerce, and enterprise IT systems operating on a 24/7 basis \cite{koomey2011}. 

This scenario highlights the urgency of developing solutions that balance operational efficiency, resilience, and sustainability, given that energy costs represent a substantial and growing portion of data center operational expenses \cite{safari2025}.
The challenge of managing energy consumption in data centers is further intensified by the increasing complexity of their infrastructure. As discussed by \cite{belady2007}, the exponential growth of data traffic and the need for intensive cooling systems significantly contribute to rising energy demands and carbon emissions \cite{flores2025}. Additionally, redundancy mechanisms, backup systems, and geographically distributed architectures—essential for ensuring reliability and availability—further increase energy consumption \cite{barroso2009}. In this context, metrics such as Power Usage Effectiveness (PUE) and Water Usage Effectiveness (WUE) have been widely adopted to evaluate operational efficiency, reinforcing the need for continuous and systematic monitoring approaches \cite{flores2025, safari2025}. 

Despite advances in monitoring and optimization techniques, a key challenge remains: how to effectively integrate real-time data collection, predictive modeling, and control mechanisms in a scalable and economically viable way, particularly for small- and medium-scale data centers \cite{huang2025,kahil2025}.While digital twin technologies have emerged as promising tools for simulating and optimizing complex systems, most existing approaches focus on large-scale infrastructures or require significant computational and financial resources, limiting their applicability in constrained environments \cite{alzami2024,pan2025}.Furthermore, the integration of Machine Learning techniques into digital twin architectures introduces additional challenges related to data quality, model generalization, and real-time processing constraints \cite{li2025,dash2025}.
This gap motivates the following research question: how effective are simplified digital twin architectures in improving energy efficiency in data centers under resource-constrained environments\cite{ba2025}?
To address this question, this paper proposes a simplified digital twin architecture that integrates IoT-based data acquisition, cloud-based processing, and Machine Learning techniques to monitor, predict, and optimize energy consumption in data center environments \cite{pan2025,gupta2023}.The proposed approach emphasizes scalability, cost-efficiency, and practical deployment using containerized services and managed cloud infrastructure \cite{furnadzhiev2025,borra2024}. The results provide empirical evidence of the feasibility of the approach, demonstrating measurable improvements in energy efficiency and operational performance metrics \cite{xu2025,huang2025}.

% ------------------------------------------------

\section{Methods}

This study adopts an applied and experimental research approach, aiming to investigate the effectiveness of a simplified digital twin architecture for energy optimization in data center environments. The research is characterized as exploratory and descriptive, as it examines the feasibility of integrating real-time data collection, cloud-based processing, and Machine Learning techniques within a constrained infrastructure context \cite{ba2025}. In line with data-driven and experimental approaches commonly adopted in complex system analysis, the methodological design is grounded in empirical observation, data-driven modeling, and experimental validation, considering the dynamic nature of data center operations \cite{kahil2025,huang2025}.

Initially, a structured literature review was conducted following PRISMA guidelines, covering major scientific databases such as IEEE Xplore, Scopus, and Web of Science. The review focused on topics including digital twins, energy efficiency, cloud computing, and containerized architectures, allowing the identification of key variables, performance metrics, and existing limitations in current approaches \cite{alzami2024,pan2025,wu2025}.This stage provided the theoretical foundation for defining the experimental design and selecting appropriate evaluation metrics, such as Power Usage Effectiveness (PUE) and predictive accuracy indicators \cite{safari2025}. 

The proposed architecture was implemented using a combination of containerized microservices and cloud-based infrastructure. Docker was employed to ensure portability and reproducibility, while the Google Cloud Platform (GCP) was used to orchestrate services such as Kubernetes Engine, Pub/Sub, and monitoring tools, enabling scalable data ingestion and processing \cite{furnadzhiev2025,borra2024}. The Machine Learning layer was developed in Python, utilizing libraries such as NumPy, Pandas, Scikit-learn, TensorFlow, and Prophet, supporting both predictive modeling and anomaly detection tasks \cite{li2025}.

For empirical validation, a controlled experimental environment was designed to simulate a small-scale data center. The testbed consisted of Dell PowerEdge R440 servers, network switches, and IoT-based sensors capable of measuring electrical and environmental variables, including power consumption, temperature, and humidity. Data were collected at a frequency of 1 Hz and transmitted via lightweight messaging protocols (MQTT) to a gateway device, which forwarded the information to the cloud infrastructure for processing and storage \cite{gupta2023}. Experimental conditions were defined by varying CPU workloads from 10\% to 90\%,allowing the analysis of the relationship between system utilization, thermal behavior, and energy consumption
\cite{xu2025}.The evaluation of the proposed approach was conducted using performance metrics designed to assess both predictive accuracy and energy efficiency. Predictive models were evaluated using Mean Absolute Error (MAE) and Root Mean Squared Error (RMSE), providing a quantitative assessment of forecasting performance \cite{li2025}.

Additionally, energy efficiency was analyzed through variations in total energy consumption (kWh) and PUE values before and after the implementation of the digital twin-based control mechanisms \cite{wu2025}. The analysis focused on identifying consistent patterns of improvement under controlled experimental conditions, supporting the assessment of the practical impact of the proposed approach \cite{flores2025,safari2025}.
By integrating real-time data acquisition, predictive modeling, and controlled experimentation, the methodology enables a comprehensive evaluation of the proposed system, ensuring both reproducibility and analytical rigor. This approach allows not only the validation of the digital twin architecture but also the assessment of its practical impact on energy efficiency in resource-constrained data center environments \cite{ba2025,pan2025}.

\subsection{Proposed Architecture}

The proposed digital twin architecture was designed to support real-time monitoring, prediction, and optimization of energy consumption in data center environments, with an emphasis on scalability, modularity, and cost-efficiency \cite{semeraro2025, ba2025}. Inspired by the conceptual framework of digital twins described in the literature, the system (shown in Figure 1) is structured into three logical layers: data acquisition (edge layer), cloud-based processing (cloud layer), and visualization and control (presentation layer) \cite{pan2025,semeraro2025}.

\begin{figure*}[t]
\centering
\includegraphics[width=\textwidth]{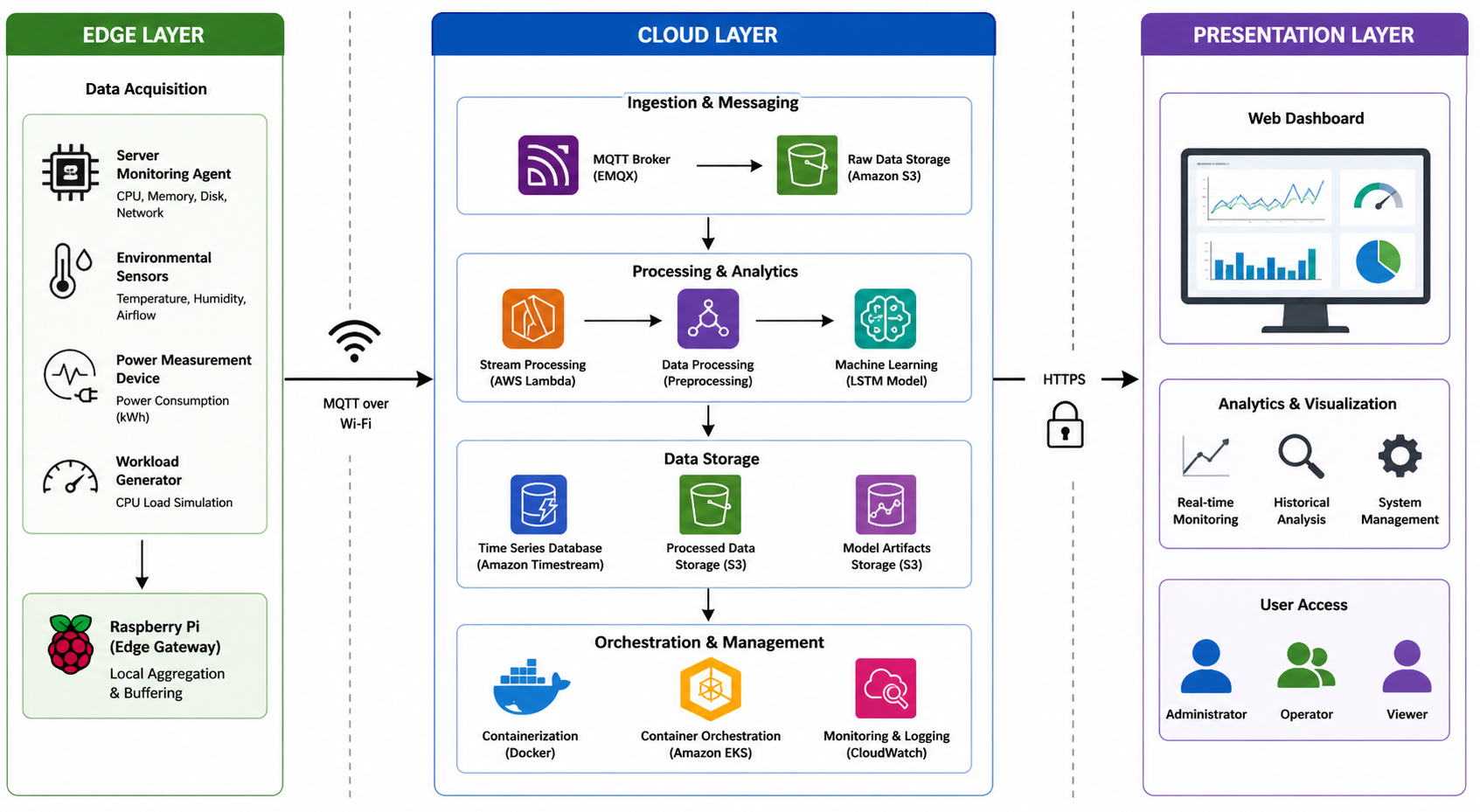}
\caption{Proposed Digital Twin Architecture}
\end{figure*}

The edge layer is responsible for data collection and initial preprocessing. It consists of IoT-based sensors deployed within the data center environment, capturing electrical and environmental variables such as power consumption, temperature, and humidity. These sensors communicate with a gateway device, which aggregates and normalizes the collected data before transmitting it to the cloud infrastructure using lightweight communication protocols \cite{gupta2023}. This design enables continuous monitoring while maintaining low computational overhead at the edge level.

The cloud layer performs data ingestion, storage, and analytical processing. It is implemented using a containerized microservices architecture deployed on a cloud platform, allowing scalability and dynamic resource allocation according to workload demands \cite{furnadzhiev2025}. Within this layer, Machine Learning models are applied to perform anomaly detection and energy consumption forecasting, enabling predictive insights that support operational decision-making \cite{li2025}. The use of asynchronous communication mechanisms ensures robustness and flexibility in handling real-time data streams.

The presentation layer provides an interface for visualization and system interaction. It includes dashboards that display real-time metrics, historical trends, and predictive outputs, allowing operators to monitor system performance and identify optimization opportunities. Additionally, this layer supports control actions, such as adjusting cooling parameters or deactivating underutilized resources, based on the insights generated by the analytical models \cite{flores2025}. Logging and monitoring mechanisms are also incorporated to ensure traceability and system reliability.

The integration of these three layers enables the creation of a functional digital twin capable of replicating key aspects of the physical system, allowing the evaluation of optimization strategies in a virtual environment before their application in real-world operations \cite{alam2023,alzami2024}. This architectural design prioritizes flexibility and reproducibility, making it suitable for deployment in small- and medium-scale data center environments where resource constraints limit the adoption of more complex solutions \cite{ba2025}. 

\subsection{Data Collection and Experimental Setup}

The empirical evaluation of the proposed digital twin architecture was conducted in a controlled experimental environment designed to simulate a small-scale data center scenario. The objective was to analyze the relationship between system workload, environmental conditions, and energy consumption under varying operational states \cite {huang2025}.
The experimental setup consisted (Table 1) of a set of physical computing resources, including enterprise-grade servers and network infrastructure, combined with IoT-based sensors capable of capturing both electrical and environmental variables. 

\begin{table}[h]
\centering
\caption{Experimental environment specifications}
\label{tab:setup}
\begin{tabular}{p{0.35\columnwidth} p{0.55\columnwidth}}
\hline
Component & Specification \\
\hline
Servers & Dell PowerEdge R440 (2× Xeon Silver, 128 GB RAM) \\
Network Switch & PoE Layer 3 \\
Cooling System & 12,000 BTU air conditioner \\
Current Sensors & True RMS, class 0.5 \\
Temperature Sensors & High-precision NTC thermistors \\
Humidity Sensors & Digital hygrometers \\
Gateway & Raspberry Pi 4 \\
Communication Protocols & MQTT / SNMP v3 / Modbus TCP \\
Sampling Frequency & 1 Hz \\
Data Storage & Google BigQuery \\
Logging Interval & 5 minutes \\
Workload Variation & 10\% to 90\% CPU (stress-ng) \\
\hline
\end{tabular}
\end{table}

The monitored parameters included power consumption, temperature, and humidity, which are widely recognized as critical factors influencing energy efficiency in data center operations \cite{xu2025,safari2025}. These variables were continuously recorded to enable the analysis of temporal patterns and system behavior under different load conditions.
Data acquisition was performed at a sampling frequency of 1 Hz, ensuring high temporal resolution for monitoring dynamic changes in system performance. The collected data were transmitted via lightweight messaging protocols to a gateway device, which handled initial data aggregation and forwarding to the cloud-based processing layer \cite{gupta2023}. This approach enabled reliable and scalable data ingestion while maintaining low latency and minimal processing overhead at the edge level.

To simulate realistic operational conditions, controlled experiments were conducted by varying CPU workloads between 10\% and 90\% over predefined time intervals. This variation allowed the observation of how changes in computational demand affect thermal behavior and overall energy consumption. The experimental design ensured consistency across test cycles, enabling comparative analysis between different system states \cite{huang2025}.

The collected data were stored in a cloud-based environment, allowing further processing and analysis by the Machine Learning models described in the subsequent sections. Data preprocessing steps included validation, normalization, and handling of missing values to ensure data quality and reliability for predictive modeling \cite{li2025}.
This experimental configuration enables the systematic evaluation of the proposed digital twin architecture, providing a structured basis for analyzing the impact of predictive and control mechanisms on energy efficiency. By combining real-time data acquisition with controlled workload variation, the methodology supports reproducibility and allows the investigation of system behavior under different operational scenarios \cite{ba2025,pan2025}. 

\subsection{Machine Learning Models}

The Machine Learning component of the proposed digital twin architecture was designed to support both energy consumption forecasting and anomaly detection, enabling data-driven insights for operational optimization  \cite{li2025}. The modeling strategy prioritized a minimal and representative set of algorithms, balancing predictive capability and computational efficiency within resource-constrained environments \cite{kahil2025} 
For the prediction task, a Long Short-Term Memory (LSTM) network was adopted as the primary model due to its ability to capture temporal dependencies in time-series data, which are inherent to energy consumption patterns in data center operations \cite{li2025}. 

As a baseline for comparison, a linear regression model was also implemented, providing a reference for evaluating the benefits of incorporating non-linear and temporal modeling approaches \cite{kahil2025}.
The models were trained using historical data collected over a 30-day period, including features such as power consumption, temperature, humidity, and system workload indicators. Standard preprocessing techniques were applied, including normalization and handling of missing values, ensuring data consistency and stability during training and evaluation \cite{li2025}.

For anomaly detection, the Isolation Forest algorithm was employed as an unsupervised method capable of identifying deviations from normal operating behavior, such as abrupt variations in energy consumption or environmental conditions \cite{dash2025}. This approach enables the detection of inefficiencies and potential system faults without requiring labeled data, making it suitable for real-time monitoring scenarios \cite{dash2025}.
Model performance was evaluated using standard regression metrics, including Mean Absolute Error (MAE) and Mean Absolute Percentage Error (MAPE), allowing the comparison between the baseline and the LSTM model across different forecasting horizons \cite{li2025}. The results demonstrate that the LSTM model provides improved predictive accuracy, reinforcing its suitability for modeling dynamic energy consumption patterns in data center environments \cite{kahil2025}.

\subsection{Evaluation Metrics}
The evaluation of the proposed digital twin architecture was conducted using a set of performance and efficiency metrics designed to assess both predictive accuracy and energy optimization outcomes \cite{li2025}.These metrics were selected to provide a comprehensive analysis of the system’s effectiveness in real-time data center environments \cite{kahil2025}.
For the predictive modeling component, standard regression metrics were employed, including Mean Absolute Error (MAE) and Mean Absolute Percentage Error (MAPE). MAE provides an absolute measure of prediction error, while MAPE enables the interpretation of errors in relative terms, facilitating comparison across different time horizons and operational conditions \cite{li2025}. These  metrics were used to evaluate the performance of the LSTM model against the baseline linear regression model.

In terms of energy efficiency, the system was evaluated based on variations in total energy consumption (kWh) and improvements in Power Usage Effectiveness (PUE), a widely adopted metric that relates total facility energy consumption to the energy consumed by IT equipment \cite{belady2007,safari2025}. Changes in these indicators were analyzed before and after the implementation of the digital twin-based monitoring and control mechanisms.

Additionally, the evaluation focused on the consistency of observed improvements in energy consumption and system performance across controlled experimental conditions, rather than formal statistical inference \cite{ba2025}. This approach emphasizes reproducibility and practical relevance, ensuring that the reported results reflect stable and observable system behavior.
Together, these metrics provide a multi-dimensional evaluation framework, enabling the assessment of both predictive accuracy and practical impact, and supporting a rigorous analysis of the proposed approach in resource-constrained data center environments \cite{pan2025}.  
% ------------------------------------------------

\section{Results}

The results obtained from the experimental evaluation of the proposed digital twin architecture are presented and analyzed in this section, with the objective of assessing its effectiveness in improving energy efficiency and predictive performance under controlled data center conditions \cite{huang2025,xu2025}. The analysis highlights key quantitative outcomes, such as reductions in energy consumption and improvements in predictive accuracy, and discusses their implications in the context of existing research on energy optimization and digital twin applications \cite {ba2025,pan2025}. The results are based on controlled experimental conditions and partially simulated data, which limits direct generalization but supports the evaluation of system feasibility and practical applicability \cite{kahil2025}. 

\subsection{Energy Consumption Reduction}

The implementation of the proposed digital twin architecture resulted in measurable improvements in energy consumption under controlled experimental conditions. A consistent reduction of approximately 10\% to 10.4\% in total energy consumption was observed when comparing baseline operation with the optimized scenario supported by the digital twin system \cite{xu2025, flores2025}.

This reduction can be primarily attributed to the identification and deactivation of underutilized computational resources, as well as dynamic adjustments in cooling parameters. The continuous monitoring of key variables, including CPU utilization, ambient temperature, and humidity, enabled the detection of inefficiencies and supported targeted interventions aimed at reducing unnecessary energy expenditure \cite{huang2025, safari2025}.

Furthermore, the integration of real-time data acquisition with predictive modeling contributed to the stabilization of system behavior, preventing demand peaks and enabling more efficient resource allocation. These findings suggest that combining monitoring with Machine Learning-based recommendations can provide practical benefits in energy management, even in constrained environments \cite{li2025, kahil2025}.
It is important to note that these results were obtained in a small-scale and partially simulated environment, which limits their direct generalization to large-scale data centers. However, the observed reduction indicates the feasibility of the proposed approach and highlights its potential for replication and further validation in more complex operational scenarios \cite{ba2025, semeraro2025}.

Overall, the results reinforce the relevance of proactive energy management strategies in data centers, particularly those based on continuous monitoring and adaptive control mechanisms. This is consistent with prior studies that emphasize the importance of workload optimization and thermal management as key factors in reducing energy consumption \cite{belady2007, barroso2009}.

\subsection{Predictive Performance}

The predictive performance of the proposed digital twin architecture was evaluated by comparing the accuracy of the selected Machine Learning models across different forecasting horizons. As presented in Table 2, the Long Short-Term Memory (LSTM) model consistently outperformed the baseline linear regression model, achieving lower prediction errors in all evaluated scenarios \cite{li2025, kahil2025}.
\begin{table}[h]
\centering
\caption{Prediction error (MAPE) across different forecasting horizons}
\label{tab:mape}
\begin{tabular}{lcc}
\hline
Forecast Horizon & LSTM (\%) & Linear Regression (\%) \\
\hline
1 hour  & 4.0  & 6.2 \\
6 hours & 7.8  & 11.3 \\
24 hours & 14.0 & 19.0 \\
\hline
\end{tabular}
\end{table}

In short-term forecasts (1 hour), the LSTM model achieved a Mean Absolute Percentage Error (MAPE) of approximately 4\%, while the linear regression model presented higher error values. This performance gap became more pronounced in medium- and long-term horizons (6 and 24 hours), where the LSTM maintained relatively stable accuracy, whereas the baseline model exhibited a significant increase in prediction error \cite{li2025}.
The superior performance of the LSTM model can be attributed to its ability to capture temporal dependencies and non-linear patterns in time-series data, which are characteristic of energy consumption dynamics in data center environments. These findings are consistent with previous studies that highlight the effectiveness of recurrent neural networks in modeling complex temporal behaviors  \cite{li2025, kahil2025, borra2024}. 

From an operational perspective, improved predictive accuracy directly enhances the system’s ability to anticipate energy demand and support proactive decision-making. More accurate forecasts allow for better scheduling of computational workloads and more efficient control of cooling systems, reducing unnecessary energy consumption and improving overall system performance \cite{huang2025,xu2025}.It is important to note that the evaluation was conducted using data collected in a controlled and partially simulated environment, which may limit the generalization of the results to large-scale deployments. However, the consistent performance gains observed across different time horizons reinforce the suitability of the proposed approach for predictive energy management in constrained data center scenarios \cite{ba2025}.                                                   
\subsection{PUE Analysis}
The overall energy efficiency of the experimental environment was evaluated using the Power Usage Effectiveness (PUE) metric, which relates total facility energy consumption to the energy consumed by IT equipment. This metric is widely adopted in the data center industry as a standard indicator of operational efficiency \cite{belady2007,safari2025}.

Prior to the implementation of the digital twin architecture, the average PUE value observed in the experimental environment was approximately 1.85. Following the adoption of the proposed monitoring and optimization mechanisms, the PUE decreased to approximately 1.70, indicating a measurable improvement in overall energy efficiency under controlled conditions \cite{flores2025, xu2025}.

\begin{figure*}[t]
\centering
\includegraphics[width=\textwidth]{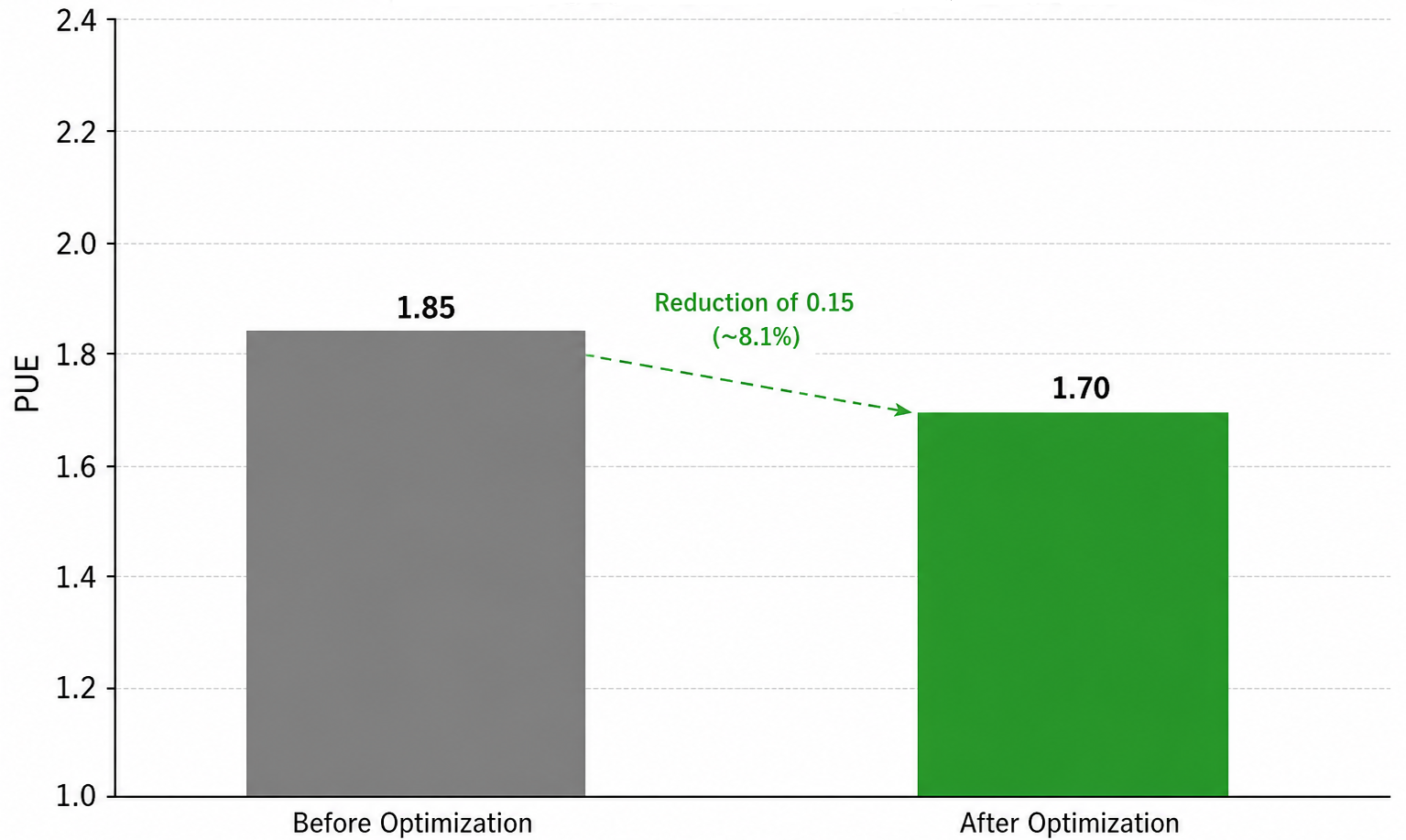}
\caption{PUE Before and After Optimization}
\end{figure*}

This improvement can be attributed to a combination of factors, including the identification and deactivation of underutilized computational resources, as well as the optimization of airflow and cooling configurations. The continuous monitoring of environmental variables enabled more precise control over thermal conditions, reducing inefficiencies associated with overcooling and suboptimal air circulation \cite{huang2025,safari2025}.Although the experimental setup represents a small-scale and partially simulated environment, the observed reduction in PUE demonstrates the potential effectiveness of digital twin-based approaches in improving energy efficiency. These findings are consistent with prior research, which highlights cooling systems and workload management as key drivers of energy optimization in data centers \cite{belady2007,barroso2009}.

From a practical perspective, even modest improvements in PUE can lead to significant energy savings when scaled to larger environments. Therefore, the results suggest that the proposed approach may provide a viable pathway for enhancing energy efficiency, particularly in small- and medium-scale data centers where cost-effective solutions are required \cite{ba2025, pan2025}.

\subsection{Cost Analysis}

The economic feasibility of the proposed digital twin architecture was evaluated by analyzing the operational costs associated with its deployment in a cloud-based environment. The results indicate that the monthly expenses related to the use of cloud infrastructure ranged between approximately USD 100 and USD 120, covering compute resources, data storage, container orchestration, and network usage \cite{borra2024,alam2023}.

These costs were influenced by the dynamic allocation of resources enabled by the containerized architecture and the use of managed cloud services. The implementation of automatic scaling mechanisms allowed the system to adjust computational capacity according to workload demand, reducing resource utilization during periods of low activity and preventing unnecessary expenses \cite{furnadzhiev2025}.
From an operational perspective, the elasticity provided by the cloud environment plays a crucial role in maintaining cost efficiency, particularly in experimental and small-scale deployments. This flexibility enables the system to balance performance and cost, making the proposed approach suitable for scenarios where resource constraints are a significant factor \cite{barroso2009,borra2024}.

Although the presented cost values are specific to the experimental setup and cloud provider configuration, they provide an indication of the economic viability of implementing digital twin architectures using commercially available infrastructure. These findings suggest that, in addition to improving energy efficiency, the proposed approach can be deployed with relatively low operational cost, which is an important consideration for adoption in academic environments and small- to medium-sized organizations \cite {ba2025,pan2025}.

% ------------------------------------------------

\section{Discussion}

The results obtained in this study provide consistent evidence that simplified digital twin architectures can contribute to improving energy efficiency in data center environments, even under constrained conditions. By integrating real-time monitoring, predictive modeling, and adaptive control mechanisms, the proposed approach demonstrated measurable improvements across multiple dimensions, including energy consumption, predictive accuracy, and operational efficiency \cite{ba2025,pan2025}.

The observed reduction of approximately 10\% in energy consumption aligns with prior research that emphasizes the role of workload optimization and dynamic resource  management in reducing energy demand in data centers \cite{belady2007, barroso2009}. In this context, the results reinforce the importance of combining monitoring infrastructures with data-driven decision-making processes, enabling more efficient allocation of computational and cooling resources.

Regarding predictive performance, the superior results achieved by the LSTM model are consistent with findings in the literature that highlight the effectiveness of recurrent neural networks in capturing temporal dependencies in energy consumption patterns \cite{li2025,kahil2025}. The ability to accurately forecast short- and medium-term demand provides a critical foundation for proactive system management, allowing operators to anticipate fluctuations and implement preventive actions. This supports the broader argument that predictive analytics is a key enabler for intelligent infrastructure management \cite{dash2025}.

The improvement observed in PUE values further corroborates the impact of the proposed approach on overall system efficiency. The reduction from 1.85 to 1.70, although obtained in a controlled and small-scale environment, demonstrates the potential of digital twin-based strategies to enhance thermal management and reduce energy waste. This finding is particularly relevant given that cooling systems are widely recognized as one of the main contributors to energy consumption in data centers \cite{belady2007}.From an economic perspective, the relatively low operational cost observed in the cloud-based implementation suggests that the proposed architecture can be deployed without significant financial barriers. This is especially important for small- and medium-scale data centers, where resource constraints often limit the adoption of advanced optimization techniques. The results indicate that combining containerization, cloud elasticity, and Machine Learning can offer a cost-effective pathway toward more sustainable infrastructure management \cite{furnadzhiev2025,borra2024}. 

Despite these promising results, it is important to acknowledge the limitations of the study. The experiments were conducted in a controlled environment with partial reliance on simulated data, which may affect the generalization of the findings to large-scale or highly heterogeneous data center environments. Additionally, the scope of the evaluation was restricted to specific workload patterns and infrastructure configurations, suggesting that further validation in diverse real-world scenarios is necessary \cite{ba2025}.

Overall, the findings contribute to the ongoing discussion on the practical applicability of digital twin technologies in energy management. By demonstrating that a simplified and resource-efficient implementation can produce measurable benefits, this study extends existing research and highlights the potential of digital twins as accessible tools for improving operational efficiency. Future work should explore the integration of renewable energy sources, more advanced predictive models, and large-scale validation to further enhance the robustness and applicability of the proposed approach \cite{pan2025,alzami2024}.

% ------------------------------------------------

\section{Conclusion}

This study investigated the feasibility and effectiveness of a simplified digital twin architecture for energy optimization in data center environments under resource-constrained conditions. The results demonstrated that the integration of real-time monitoring, predictive modeling, and adaptive control mechanisms can lead to measurable improvements in energy efficiency, including reductions in total energy consumption and improvements in PUE values \cite{ba2025,flores2025}.

The findings also highlighted the relevance of Machine Learning techniques, particularly LSTM models, in supporting predictive energy management by capturing temporal patterns and enabling proactive decision-making \cite{li2025,kahil2025}. Additionally, the use of containerized cloud-based infrastructure proved to be a viable and cost-effective approach, reinforcing the practical applicability of the proposed solution in small- and medium-scale environments \cite{furnadzhiev2025, borra2024}. 

Despite these contributions, the study is subject to limitations, including the use of controlled experimental conditions and partially simulated data, which may restrict the generalization of the results. Therefore, further research is needed to validate the proposed architecture in large-scale and heterogeneous data center environments, as well as to explore the integration of additional optimization strategies and energy sources \cite{ba2025}.
Overall, this work contributes to the field by demonstrating that simplified and resource-efficient digital twin implementations can provide tangible benefits in energy optimization, offering a practical pathway for improving the sustainability and operational efficiency of data center infrastructures \cite {pan2025,alzami2024}.

% ------------------------------------------------
% ACKNOWLEDGMENTS
% ------------------------------------------------

\section*{Acknowledgments}

The authors would like to thank the Federal Institute of Education, Science and Technology of São Paulo (IFSP) for institutional support.

% ------------------------------------------------
% REFERENCES
% ------------------------------------------------

\bibliographystyle{plain}
\bibliography{sample}

\end{document}